\newcommand{\be}{\begin{equation}}
\newcommand{\ee}{\end{equation}}
\newcommand{\bea}{\begin{eqnarray}}
\newcommand{\eea}{\end{eqnarray}}
\newcommand{\nn}{\nonumber}
\newcommand{\Appendix}[1]%
    {\renewcommand{\thesection}{Appendix~\Alph{section}:}%
         \section{#1}}%
\catcode`@=11
\long\def\@makecaption#1#2{
   \vskip 10pt
   \setbox\@tempboxa\hbox{{\small\bf #1.} \ {\small #2}}
   \ifdim \wd\@tempboxa >\hsize       % IF longer than one line:
   {\small\bf #1.} \ {\small #2}\par  % THEN set as ordinary paragraph.
   \else                              %   ELSE  center.
        \hbox to\hsize{\hfil\box\@tempboxa\hfil}
   \fi}
\catcode`@=12

\catcode`@=11
\def\secteqno{\@addtoreset{equation}{section}%
\def\theequation{\thesection.\arabic{equation}}}
\def\endsecteqno{\def\theequation{\@ifundefined{chapter}%
{\arabic{equation}}{\thechapter.\arabic{equation}}}}
\newcounter{subequation}
\def\thesubequation{\alph{subequation}}
\def\sneqnarray{\stepcounter{equation}\let\@currentlabel=\theequation
\setcounter{subequation}{1}
\def\@eqnnum{{\rm (\theequation\thesubequation)}}
\global\@eqcnt\z@\tabskip\@centering\let\\=\@eqncr\let\@@eqncr=\@@sneqncr
$$\halign to \displaywidth\bgroup\@eqnsel\hskip\@centering
 $\displaystyle\tabskip\z@{##}$&\global\@eqcnt\@ne
 \hskip 2\arraycolsep \hfil${##}$\hfil
 &\global\@eqcnt\tw@ \hskip 2\arraycolsep
$\displaystyle\tabskip\z@{##}$\hfil
  \tabskip\@centering&\llap{##}\tabskip\z@\cr}
\def\endsneqnarray{\@@sneqncr\egroup $$\global\@ignoretrue}
\def\@@sneqncr{\let\@tempa\relax
   \ifcase\@eqcnt \def\@tempa{& & &}\or \def\@tempa{& &}
   \else \def\@tempa{&}\fi
     \@tempa \if@eqnsw\@eqnnum\stepcounter{subequation}\fi
     \global\@eqnswtrue\global\@eqcnt\z@\cr}
\def\nobiblabels{\def\@lbibitem[##1]##2{\@bibitem{##2}}}
\catcode`@=12

 \def\pa{\partial}  \def\dag{\dagger}
 \def\nb{\bm{\nabla}}  \def\nbg{\overleftrightarrow{\bm{\nabla}}}
\def\cbg{\overleftrightarrow{\bm{D}}}
\documentclass[a4paper,11pt,preprint,aps,prd,showpacs,amsmath,amssymb,nofootinbib,showkeys,superscriptaddress]{revtex4-1}

\usepackage[german,english]{babel}
\usepackage{hyperref}
\usepackage{amssymb}
\usepackage{amsmath}
\usepackage{bm}% bold math   
\usepackage{braket}
\usepackage{slashed}
\usepackage{multirow}
\usepackage{epsfig}
\usepackage{epstopdf}
\usepackage{color}
\usepackage{soul}
\usepackage{bbm}

\begin{document}

\preprint{TUM-EFT 108/18}

\title{Effective field theory for the nucleon-quarkonium interaction}

\author{Jaume Tarr\'us Castell\`a}
\email{jtarrus@ifae.es}
\affiliation{Grup de F\'\i sica Te\`orica, Dept. F\'\i sica and IFAE-BIST, 
Universitat Aut\`onoma de Barcelona, \\ E-08193 Bellaterra, Catalonia, Spain}
\author{Gast\~ao Krein}
\email{gkrein@ift.unesp.br }
\affiliation{Instituto de F\'isica Te\'orica, Universidade Estadual Paulista, \\
Rua Dr.~Bento Teobaldo Ferraz, 271 - Bloco II, 01140-070 S\~ao Paulo, SP, Brazil}

\date{\today}
\begin{abstract}
We develop an effective field theory approach for the $S$-wave quarkonium-nucleon system. 
We adopt a natural power counting equivalent to Weinberg's power counting in 
nucleon-nucleon effective field theories and compute the quarkonium-nucleon potential, 
scattering length and effective range up to $\mathcal{O}(m^3_{\pi}/\Lambda^3_{\chi})$ 
accuracy, including the full light-quark mass dependence. We compare our results with 
lattice QCD studies of quarkonium-nucleon system, obtain an estimation of the leading order quarkonium-nucleon 
contact term and determine the $J/\psi$ and $\eta_c$ chromopolarizabilities.
\end{abstract}


\maketitle

\section{Introduction}

The confirmation~\cite{Aaij:2016phn} by the LHCb collaboration at the CERN Large Hadron Collider~(LHC) 
of its earlier finding~\cite{Aaij:2015tga} of the pentaquark states $P^+_c(4380)$ and $P^+_c(4450)$ 
has sparked off renewed interest in the low-energy quarkonium-nucleon system. 
These pentaquark states appear as intermediate resonant states in the weak decay process $\Lambda^0_b 
\rightarrow J/\Psi p K^-$, have valence quark content $P^+_c = \bar{c}cuud$, and lie close 
to several charmonium-baryon thresholds~\cite{Burns:2015dwa}. Like with the plethora of X,Y,Z exotic
hadrons~\cite{Briceno:2015rlt,Lebed:2016hpi}, the underlying QCD dynamics governing the 
internal structure of these pentaquark states is not understood {\textemdash} see 
Ref.~\cite{Ali:2017jda} for a recent review.

Early interest in low-energy quarkonium-nucleon interaction was motivated by the 
fact that it can probe the mass distribution inside a nucleon. The amplitude for quarkonium-nucleon 
forward scattering at low energies has been described~\cite{Kaidalov:1992hd,Luke:1992tm,
Kharzeev:1995ij,deTeramond:1997ny,Kharzeev:1998bz,Sibirtsev:2005ex} as a product of the 
quarkonium-gluon interaction, parametrized in terms of the quarkonium 
chromopolarizability ~\cite{Peskin:1979va,Bhanot:1979vb,Voloshin:1979uv,Leutwyler:1980tn,Fujii:1999xn,
Brambilla:2015rqa}, and a matrix element of gluon fields inside the nucleon that can be obtained by 
using the trace anomaly, a quantum anomaly in the trace of the QCD energy-momentum 
tensor~\cite{Chanowitz:1972vd,Crewther:1972kn,Voloshin:1980zf}. 

The interest in quarkonium-nucleon systems is further motivated 
by upcoming experiments at the Facility for Ion Research (FAIR) aiming at the determination of cross sections 
of quarkonium propagation in nuclear matter, which are crucial for disentangling cold matter from 
deconfinement effects in experiments of relativistic heavy-ion collisions at the Relativistic Heavy 
Ion Collider (RHIC) and the LHC~\cite{Brambilla:2014jmp}. 

Quarkonium-nucleon systems have no valence quarks in common and their dynamics appears to be dominated 
by multigluon van der Waals interactions~\cite{Brodsky:1989jd,Luke:1992tm,Brodsky:1997gh,Beane:2014sda}.  
These interactions are not expected to be repulsive at short distances, a feature that
led to the interesting conjecture~\cite{Brodsky:1989jd} that quarkonia states, like the charmonia 
$J/\Psi$ and $\eta_c$, can form exotic nuclear bound states. Since this earlier conjecture, many 
different methods have been used to investigate the possible existence of such exotic states and a 
large literature on the subject has accumulated along the years {\textemdash} for a recent review, see 
Ref.~\cite{Krein:2017usp}. There is scarce experimental information on quarkonium-nucleon interaction
at low energies and practically all of our knowledge on the interaction comes from lattice 
simulations~\cite{Yokokawa:2006td,Liu:2008rza,Kawanai:2010ru,Kawanai:2010ev,Kawanai:2011zz,Alberti:2016dru,
Sugiura:2017vks}, although new, but preliminary experimental results by the GlueX collaboration 
at the Thomas Jefferson National Accelerator Facility (JLAB) have been communicated recently~\cite{stevens}. 
The lattice results, obtained for fairly large pion masses, find that indeed the charmonium-nucleon 
interaction does not contain a repulsive core, being attractive but not very strong. Another 
recent lattice QCD simulation~\cite{Beane:2014sda} found that charmonia can form nuclear bound states, with large binding energies, much larger than phenomenological expectations~\cite{Krein:2017usp,Tsushima:2011kh}, a feature that possibly is due 
to the very large pion mass used in that simulation. Regarding charmonium-nucleon bound states, while the lattice 
simulation of Ref.~\cite{Beane:2014sda} finds deeply bound nuclear states, a more recent 
simulation~\cite{Alberti:2016dru} using a smaller pion mass, finds very small binding, of the order 
of the deuteron binding energy.

The development of a theoretical framework built on controllable approximations that can be systematically 
improved is essential for the understanding of the quarkonium-nucleon system. 
Van der Waals forces have been studied in an effective field theory (EFT) framework for QED in 
Ref.~\cite{Brambilla:2017ffe} and for quarkonium-quarkonium systems in Ref.~\cite{Brambilla:2015rqa}. 
In the present paper we aim at the construction of a nonrelativistic chiral EFT for the $J/\psi$ and 
$\eta_c$ interactions with nucleons, which we name quarkonium-nucleon EFT, QNEFT. We restrict the 
discussion to $S$-wave quarkonia but higher states can be treated similarly. QNEFT is 
constructed by coupling an $S$-wave quarkonium field with a nucleon doublet and a pion triplet 
according to chiral symmetry, heavy-quark spin symmetry, and CPT symmetry. We adopt a counting 
scheme analogous to Weinberg's counting for nucleon-nucleon EFT~\cite{Weinberg:1991um,Weinberg:1990rz},
that is, the natural power counting based on dimensional analysis. The nucleon-quarkonium dynamics takes 
place at a lower energy scale than the pion mass. Therefore, it is convenient to integrate out the dynamics 
at the $E\sim m_{\pi}$ scale and match QNEFT to a lower energy EFT, which we call potential 
quarkonium-nucleon EFT, pQNEFT, in which the $S$-wave quarkonium interacts through contact and 
potential interactions. We have put special emphasis on obtaining the light-quark mass dependence 
of all the matching coefficients. Within pQNEFT we obtain the expressions for the scattering length 
and effective range as a function of the light-quark mass. We then compare with studies on the lattice 
of the potential and effective range expansion (ERE) parameters which allows us to test our assumptions 
on the counting and determine the $J/\psi$ and $\eta_c$ chromopolarizabilities. Finally, the EFT framework 
allows us to discuss the compatibility of the results from the different lattice simulations.

The paper is organized as follows. In Sec.~\ref{nqeft} we detail the nonrelativistic EFT for nucleon 
and $S$-wave quarkonium (QNEFT) interacting at energies of the order of the pion mass $E\sim m_{\pi}$. 
In Sec.~\ref{pnqeft} we introduce the lower-energy EFT, $E\sim m^2_{\pi}/\Lambda_{\chi}$, termed pQNEFT, 
and work out the matching coefficients. The scattering amplitude, scattering length and effective range 
in pQNEFT are computed up to next-to-next-to-next-to-leading order (N$^3$LO) including the full 
light-quark mass dependence in Sec.~\ref{erep}. We compare our result for the quarkonium-nucleon 
potential and the scattering length and effective range with the HAL QCD method and lattice determinations in Sec.~\ref{cnpla} 
and in Sec.~\ref{cnplb} respectively. We present a discussion of our results and conclusions in 
Sec.~\ref{conc}. In Appendix~\ref{appa} we detail the calculation of the quarkonium-nucleon potential 
in coordinate space.

\section{Quarkonium-Nucleon EFT}\label{nqeft}

We consider an EFT of QCD containing as degrees of freedom $S$-wave quarkonia, nucleons and pions 
at energies of order $m_{\pi}$, much below $\Lambda_{\chi} \sim 1~$GeV, the scale of dynamical chiral 
symmetry breaking. Since the masses of the nucleon and $S$-wave quarkonia are close or above 
$\Lambda_{\chi}$, a nonrelativistic formulation for both fields is convenient~\cite{Jenkins:1990jv}. 
In the following, we write the Lagrangian density necessary for 
nucleon-$S$-wave quarkonium scattering up to terms of $\mathcal{O}(m^3_{\pi}/\Lambda^3_{\chi})$, 
consistent with chiral symmetry, $C$ and $P$ invariance and rotational symmetry (and Lorentz 
invariance in the Goldstone boson sector).

As a basic building block we use the unitary matrix $U(x)$ to parametrize the Goldstone boson fields:
\be
u^2=U = e^{i\Phi /F}\,, \quad 
\Phi=\left( \begin{array}{cc}
\pi^0 & \sqrt{2} \pi^+\\
\sqrt{2} \pi^- & -\pi^0
\end{array}
\right)\,.
\label{pions}
\ee
At leading order, $F$ may be identified with the pion decay constant $F_\pi=92.419$~MeV. The most 
convenient choice of fields for the construction of the effective Lagrangian is given by 
$u_{\mu}=i\left\{u^{\dag},\partial_{\mu}u\right\}$ and $\chi_{\pm} = u^{\dag}\chi u^{\dag}\pm u 
\chi^{\dag} u$, where $\chi = 2B\hat{m}\mathbbm{1}$, with $B$ being related to the vacuum quark condensate
and $\hat{m}$ is the average of the $u$ and $d$ quark masses. We work in the isospin limit, 
$m_u = m_d \equiv \hat{m}$; in this limit, the pion mass is $m_{\pi}^2 = 2 B\hat{m} \approx (135$~MeV$)^2$. 
The Lagrangian density for the Goldstone boson sector at leading order is given by
\be
\mathcal{L}^{\pi}=\frac{F^2}{4}\left(\langle u_{\mu}u^{\mu}\rangle+\langle\chi_+\rangle\right)\,,
\ee
where $\langle \cdots \rangle$ means trace over flavor. We are interested only in the $S$-wave 
color-singlet quarkonia states. These can be spin singlet or spin triplet. Since the spin-dependent 
interactions are suppressed by the heavy quark mass and are beyond the accuracy we are aiming 
for, these two states can be taken as degenerate and we will represent them both with a $0^{-+}$ field 
$\phi$. The kinetic part of the $\phi$ Lagrangian density is 
\be
\mathcal{L}^{\phi} = \phi^{\dag}\left(i \partial_0+\frac{\nb^2}{2\hat{m}_{\phi}}\right)\phi\,.
\label{phil}
\ee
Since the $\phi$ field is scalar under chiral symmetry, the $\phi$-pion interactions are
described by
\begin{align}
\mathcal{L}^{\phi-\pi}&=\frac{F^2}{4}\phi^{\dag}\phi\left(c_{d0}\langle u_0u_0\rangle+c_{di}
\langle u_i u^i\rangle+c_m\langle\chi_+\rangle\right),
\label{piphil} 
\end{align}
where the low-energy constants $c_{d0}$, $c_{di}$ and $c_m$ can be determined by using the QCD trace 
anomaly~\cite{Brambilla:2015rqa} for sufficiently compact quarkonia:
\be
c_{d0} = -\frac{4\pi^2\beta}{b}\kappa_1\,, \qquad
c_{di} = -\frac{4\pi^2\beta}{b}\kappa_2\,, \qquad
c_{m}  = -\frac{12\pi^2\beta}{b}\,. \label{mcm}
\ee
with $\beta$ the chromopolarizability of the $S$-wave quarkonium state, $b$ is the first coefficient 
of the QCD beta function, $b=(11N_c-2N_f)/3$, where $N_c=3$ and $N_f=3$ are the number of colors 
and of active flavors at the quarkmonium state scale, respectively, and $\kappa_1=2-9\kappa/2$, 
$\kappa_2=2+3\kappa/2$. Here, $\kappa$ is a parameter that can be obtained from pionic transitions of quarkonium states. We will use the value 
$\kappa=0.186(9)$, extracted from $\psi^{\prime}\to J/\psi \pi^+\pi^-$ decays~\cite{Bai:1999mj}. The expressions in Eq.~\ref{mcm} Eq.~(\ref{mcm}) are valid up to corrections of $\mathcal{O}(\alpha_s)$~\cite{Novikov:1980fa} stemming from considering higher order terms in the QCD beta function appearing in the trace of the 
QCD energy-momentum tensor.

The nucleons form an isospin doublet. The nucleon sector, including couplings to pions, is described by a Lagrangian density that reads~\cite{Gasser:1987rb,Fettes:2000gb}
\be
\mathcal{L}^{N}=N^{\dag}\left(i D_0+\frac{\bm{D}^2}{2\hat{m}_N}\right)N
-\frac{g_A}{2}N^{\dag}\bm{u}\cdot\bm{\sigma}N\,.
\label{npil}
\ee 
The covariant derivative acting on the nucleon fields is defined as~\cite{Gasser:1987rb} $D_{\mu}N = \partial_{\mu}N + \Gamma_{\mu} N$, with $\Gamma_{\mu} = \frac{1}{2} \left[u^{\dag},\partial_{\mu} u\right]$.

The last pieces of the effective Lagrangian are contact terms with nucleons and $S$-wave quarkonium, which we write as
\begin{align}
\mathcal{L}^{\phi-N}=&-c_0 N^{\dag}N \phi^{\dag}\phi-d_m\langle\chi_+\rangle N^{\dag}N\phi^{\dag}\phi
-d_{1}\nb\left(N^{\dag}N\right)\cdot\nb\left(\phi^{\dag}\phi\right) \nonumber \\
&-d_{2}\left(N^{\dag}\cbg N\right)\cdot\left(\phi^{\dag}\nbg\phi\right)
-d_{3}\bm{D}N^{\dag}\cdot \bm{D}N\phi^{\dag}\phi-d_{4}N^{\dag}N\nb\phi^{\dag}\cdot \nb\phi\,,
\label{ctl}
\end{align}
with $\overleftrightarrow{\nabla} = \overleftarrow{\nabla}- \overrightarrow{\nabla}$ and analogously 
for $\overleftrightarrow{D}$. There are further next-to-leading order (NLO) terms coupling $\phi$-$N$ 
and pions but we only display the ones that do not vanish in the vacuum configuration ($u=\mathbbm{1}$).

We adopt a standard chiral counting, with $\partial_0\sim\partial_i\sim m_{\pi}$, and low-energy constants 
of $\mathcal{O}(\Lambda^{4-k}_{\chi})$ where $k$ is the dimension of the accompanying operator. Thus, the 
scaling of low-energy constants in Eq.~\eqref{ctl} is $c_0\sim1/\Lambda^2_{\chi}$, and $d_m$ as well as
$d_i,\; i=1, \cdots, 4$, scale as $d_m, d_i \sim1/\Lambda^4_{\chi}$. This power counting setup 
is  equivalent to Weinberg's power counting in nucleon-nucleon EFT. Moreover, we will count the 
quarkonium and nucleon mass as $m_N\sim m_{\phi}\sim \Lambda_{\chi}$. A more refined counting, 
taking into account $\Lambda_{\chi} \sim m_N < m_{\phi}$, is not required at the precision we aim 
for.

\section{Potential quarkonium-nucleon EFT}\label{pnqeft}

The QNEFT introduced in Sec.~\ref{nqeft} describes the interaction of nucleons and $S$-wave quarkonia 
with relative momentum $\sim m_{\pi}$. However, the quarkonium-nucleon dynamics occurs at the energy 
scale $E\sim m^2_{\pi}/\Lambda_{\chi}$. For energies $E\sim m^2_{\pi}/\Lambda_{\chi}\ll m_{\pi}$, the 
pion fields can be integrated out. This integration produces quarkonium-nucleon potentials and 
redefinitions of the low-energy constants. The Lagrangian density for pQNEFT reads as follows
\begin{align}
\mathcal{L}^{\text{pQNEFT}}=&N^{\dag}\left(i \pa_0+\frac{\nb^2}{2m_N}\right)N+\phi^{\dag}
\left(i \pa_0+\frac{\nb^2}{2m_\phi}\right)\phi-C_0 N^{\dag}N\phi^{\dag}\phi - D_{1}\nb
\left(N^{\dag}N\right)\cdot\nb\left(\phi^{\dag}\phi\right) \nonumber\\
&-D_{2}\left(N^{\dag}\nbg N\right)\cdot\left(\phi^{\dag}\nbg\phi\right)-D_{3}\nb N^{\dag}\cdot 
\nb N\phi^{\dag}\phi-D_{4}N^{\dag}N\nb\phi^{\dag}\cdot \nb\phi \nonumber\\
&-\int d^3r N^{\dag}N(t,\bm{x}_1)V(\bm{x}_1-\bm{x}_2)\phi^{\dag}\phi(t,\bm{x}_2)\,,
\end{align}
with  $\bm{r}=\bm{x}_1-\bm{x}_2$. 

The nucleon and $S$-wave quarkonium masses receive contributions from operators proportional 
to the quark masses as well as from pion loop contributions. These matching contributions can 
be found in Ref.~\cite{Procura:2003ig}  and Ref.~\cite{Brambilla:2015rqa} for the nucleon 
and $S$-wave quarkonium masses respectively. Up to contributions of $\mathcal{O}\left((m_{\pi}/
\Lambda_{\chi})^3\right)$, they are given by\footnote{The nucleon and quarkonium masses in 
Refs.~\cite{Procura:2003ig,Brambilla:2015rqa} are computed to higher accuracy, we reproduce 
only those terms that will be needed in this work.}
\begin{align}
m_N &= \hat{m}_N - 4 c_1 m^2_{\pi}-\frac{3g^2_Am^3_{\pi}}{32\pi F^2}
\,,
\label{mnq}
\\ 
m_{\phi}&=\hat{m}_{\phi}-F^2c_m m^2_{\pi} 
\,. 
\label{mphiq}
\end{align}

\begin{figure}[ht]
\centerline{\includegraphics[width=.8\textwidth]{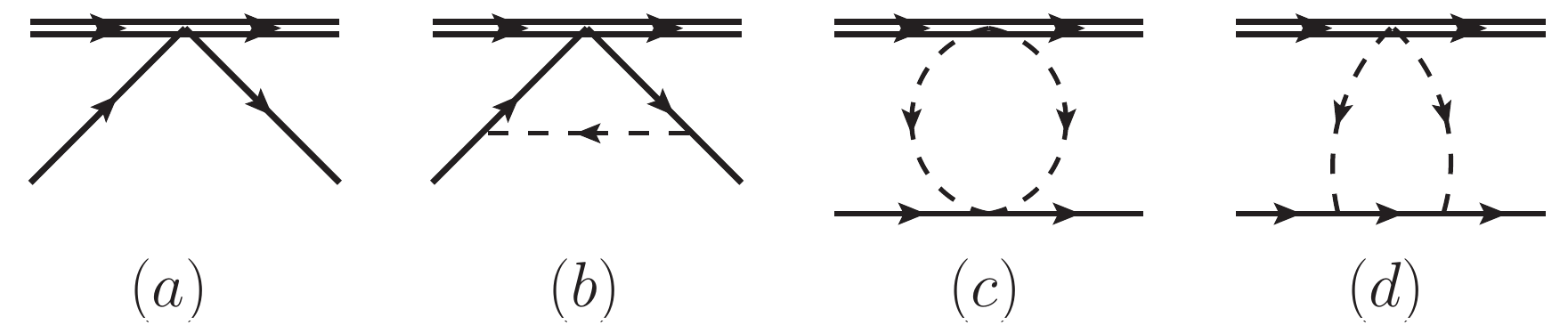}}
\caption{Diagrams contributing to nucleon-$S$-wave quarkonium scattering. Double, single and dashed lines correspond 
to quarkonium, nucleon and pions respectively.}
\label{nqsd}
\end{figure}

The matching contributions to the quarkonium-nucleon contact terms and potential up to next-to-next-to-leading
order (N$^2$LO) are shown in Fig.~\ref{nqsd}. The LO contribution is formed by the contact term proportional 
to $c_0$ from the Lagrangian \eqref{ctl}, diagram (a) in Fig.~\ref{nqsd}. The NLO contribution is given by the
operators proportional to $d_m$ and $d_i$, $i=1,...,4$, as well as the one-pion exchange between the nucleon 
legs, diagram (b) in Fig.~\ref{nqsd}. The ultraviolet divergence in diagram (b) can be renormalized 
in the $\overline{\rm MS}$ scheme by 
\be
d_m\to d_m + \frac{9g^2_Ac_0}{256\pi^2F^2}\lambda\,,\quad \lambda=2/(4-d) + 1 - \gamma_E + \log 4\pi\,,
\ee
with $\gamma_E$ being the Euler's constant. The N$^3$LO terms correspond to the two-pion exchange 
diagram in Fig.~\ref{nqsd}. Diagram (c) cancels due to the isospin structure of the vertices:
the two-pion-$\phi$ vertex requires both pions to carry the same isospin while the two-pion-nucleon 
vertex has the two pions necessarily with different isospin. 

Diagram (d) generates a potential interaction as well as local terms. We separate the local and 
nonlocal contributions of diagram (d) using a dispersive representation, the details of which can 
be found in Appendix~\ref{appa} and assign the local terms to the contact interactions. The matching 
reads: 
\begin{align}
C_0=&c_0+4m^2_{\pi}d_m 
+ \frac{9g^2_Am^2_{\pi}c_0}{64\pi^2F^2} \, \left(\log \frac{m^2_{\pi}}{\nu^2}
+\frac{2}{3}\right)
+ \frac{3g^2_Am^3_{\pi}}{64\pi F^2}\left(5c_{di}-3c_{m}\right)\,, \label{C0} 
\\
D_1=&d_1+\frac{g^2_A m_{\pi}}{256\pi F^2}\left(23c_{di}-5c_{m}\right) \,, \label{D1}
\\
D_j=&d_j \quad\text{for}\quad j=2,3\,\text{and}\,4 \,, \label{Dj}
\\
V(r)=& \frac{3g^2_A}{128\pi^2F^2 r^6} \, e^{-2m_{\pi}r}
\left\{ c_{di}
\left[6+m_{\pi}r(2+m_{\pi}r)(6+m_{\pi}r(2+m_{\pi}r))\right]\right. \nonumber\\
&\left.
+ c_m m^2_{\pi}r^2\left(1+m_{\pi}r\right)^2
\right\}\,,
\label{Vnloc}
\end{align}
with $\nu$ being the renormalization scale. Expanding the potential for long distances, 
$r \gg (2m_\pi)^{-1}$, we obtain the expression
\be
V(\bm{r}) =  
\frac{3g^2_Am^4_{\pi}\left(c_{di}+c_m\right)}{128\pi^2F^2}
\frac{e^{-2m_{\pi}r}}{r^2}\,.
\label{ldbp}
\ee
The $e^{-2m_\pi r}/r^2$ falloff was obtained previously in the large $N_c$ limit in the context 
of a chiral soliton model~\cite{Goeke:2007fp,Polyakov:2018aey}. Nevertheless, to our knowledge, 
our derivation of Eq.~\eqref{ldbp} is the first model-independent determination within a 
first-principles EFT framework.

\section{Effective range expansion parameters}
\label{erep}

In pQNEFT we can compute the $S$-wave nucleon-quarkonium scattering amplitude. For this we will need 
the contribution of the $n$-bubble diagrams for the $C_0$ contact interaction, which are given by
\be
\mathcal{A}^{n-\text{bubble}}=-C_0\left(-\frac{i\mu p\, C_0}{2\pi}\right)^n\,,
\ee
where $\mu=m_{\phi}m_N/(m_{\phi}+m_N)$ is the reduced mass of the system, and $p=\sqrt{2\mu E}$ and 
$E$ are the center of mass (c.m.) momenta and energy. The $S$-wave scattering amplitude\footnote{The partial-wave decomposition of the scattering amplitude is defined as $\mathcal{A}_l(p) = \frac{1}{2} \int^{+1}_{-1} dx \, P_l(x)\,{\mathcal{A}}(p,x)\,.$} to N$^3$LO 
reads
\begin{align}
\mathcal{A}_S^{(0)}&=-C_0\,,\label{pefta0}\\
\mathcal{A}_S^{(1)}&=-C_0\left(-\frac{i\mu p\,C_0}{2\pi}\right)\,,\\
\mathcal{A}_S^{(2)}&=-2(D_1+D_2)p^2-C_0\left(-\frac{i\mu p\,C_0}{2\pi}\right)^2\,,\\
\mathcal{A}_S^{(3)}&=-C_0\left(-\frac{i\mu p\,C_0}{2\pi}\right)^3-\widetilde{V}_S(p)
\label{pefta3}\,,
\end{align}
where the superscript $(k)$ in ${\cal A}^{(k)}_S$ indicates the suppression relative to the leading 
order amplitude in powers $(m_{\pi}/\Lambda_{\chi})^k$, and $\widetilde{V}_S(p)$ is the projection of the potential in momentum space on the $S$-wave channel:
\begin{align}
\widetilde{V}_S(p)=&\frac{3g^2_Am^3_{\pi}}{32\pi F^2}
\left\{
\left[\left(\frac{m_{\pi}}{p}+\frac{2p}{3m_{\pi}}\right)(c_{di}-c_m)
+ \left(\frac{2p}{3m_{\pi}}
+ \frac{4p^3}{5m_{\pi}^3}\right)c_{di}\right]\arctan\frac{p}{m_{\pi}}
\nonumber\right.\\ 
& \left. 
+ \frac{1}{30} \frac{m_{\pi}^2}{p^2}\left(5c_m-7c_{di}\right)\log\left(1+\frac{p^2}{m_{\pi}^2}\right)
-\frac{1}{60}\left(46+\frac{67p^2}{m^2_{\pi}}\right)c_{di}+\frac{5}{12}
\left(2+\frac{p^2}{m^2_{\pi}}\right)c_{m}
\right\}\,.\label{swpot}
\end{align}
For $\left(m_{\pi}/\Lambda_{\chi}\right)^4$ precision, we would need higher order contact terms in the QNEFT
Lagrangian in Eq.~\eqref{ctl}, with four derivatives or pion mass insertions; additional $2$-pion loop diagrams, 
analogous to the diagram (c) of Fig.~\ref{nqsd}, with NLO pion-nucleon vertices instead of LO ones; and the 
contributions to the nucleon and quarkonium masses up to the aforementioned order. Corrections to the 
pion-nucleon axial vector coupling and to higher order two-pion-quarkonium vertices contribute only starting 
at order $\left(m_{\pi}/\Lambda_{\chi}\right)^5$. Four-pion exchanges appear at most at $\left(m_{\pi}/\Lambda_{\chi}\right)^5$
precision. Two-kaon and two-$\eta$ exchanges are suppressed by two mechanisms: diagram (d) in Fig.~\ref{nqsd} is not allowed
 because the nucleon cannot emit a meson with strangeness unless it transitions into a hyperon and
diagram (c) vanishes at leading order in an analogous way as for pion exchanges. Furthermore, the two-kaon and two-$\eta$ exchange contributions are
suppressed by powers of $(m_{\pi}/m_{K})$ and $(m_{\pi}/m_{\eta})$ respectively.

The effective range expansion (ERE) parametrizes the low-energy scattering amplitudes. To fix the convention, 
let us write the on-shell S-matrix for a particular partial wave $l$ as:
\begin{align}
\mathcal{A}^{ERE}_l&=\frac{2\pi}{\mu}\frac{1}{p\cot\delta_l-ip}\label{eref}\,,
\end{align}
where $\delta_l$ is the phase shift. The ERE is defined as
\be
p^{2l+1} \cot \delta_l=-\frac{1}{a_l} + \frac{1}{2}r_l p^2+\dots \,.
\label{eredef}
\ee
We expand the $S$-wave ERE amplitude for small~$p$
\be
\mathcal{A}^{ERE}_S=-\frac{2\pi}{\mu} a_0 + \frac{2\pi}{\mu} a^2_0(i p)+\frac{2\pi}{\mu}
\left(a^3_0-\frac{1}{2}r_0a^2_0\right)p^2+\dots\,,
\ee
and match it to the pNQEFT amplitude of Eqs.~\eqref{pefta0}-\eqref{pefta3} to obtain the 
scattering length and effective range \cite{Kaplan:1998we}:
\begin{align}
a_0&=\frac{\mu}{2\pi}\left[c_0+4d_m m_{\pi}^2
+ \frac{9g^2_Am^2_{\pi}c_0}{64\pi^2F^2} \left(\log \frac{m^2_{\pi}}{\nu^2}
+\frac{2}{3}\right)+\frac{3g_A^2}{64\pi F^2}
m_{\pi}^3\left(5c_{di}-3c_m\right)\right]\,,
\label{slf}\\
r_0&=\frac{8\pi}{\mu c^2_0}\left[(d_{1}+d_{2})+\frac{g_A^2}{256\pi F^2}m_{\pi}
\left(23c_{di}-5c_m\right)\right]\label{erf}\,.
\end{align}
Note that the reduced mass is in terms of the physical quarkonium and nucleon masses, which also 
carry a dependence on $m_{\pi}$, that needs to be taken into account for chiral extrapolations. 
This dependence can be found in Eqs.~\eqref{mnq} and \eqref{mphiq} up to the accuracy required 
in Eqs.~\eqref{slf} and \eqref{erf}.

\section{Comparison with the scattering length from lattice QCD}
\label{cnplb}

\begin{table}[ht]
%\begin{ruledtabular}
\begin{tabular}{c|c|c|c|c|c}
\hline\hline
              Reference                        &                             & Channel  & $a_0$~[fm]    
              & $\,c_0~[{\rm GeV}^{-2}]\,$ & $\,d_m~[{\rm GeV}^{-2}]\,$       \\ \hline
 \multirow{4}{*}{\cite{Yokokawa:2006td}}       & \multirow{2}{*}{PSF}        & $\eta_c$ & $-0.70(66)$ 
 & $-31(29)$            & \multirow{4}{*}{\text{Quenched}} \\
                                               &                             & $J/\psi$ & $-0.71(48)$ 
                                               & $-31(21)$            &                                  
                                               \\ \cline{2-5}
                                               & \multirow{2}{*}{LLE}        & $\eta_c$ & $-0.39(14)$
                                                & $-17(6)$             &                                  
                                                \\
                                               &                             & $J/\psi$ & $-0.39(14)$ 
                                               & $-17(6)$             &                                  
                                               \
                                               \\ \hline
 \multirow{2}{*}{\cite{Kawanai:2010ru}}        &                             & $\eta_c$ & $-0.25(5)$  
 & $-8(2)$              & \multirow{2}{*}{\text{Quenched}} \\
                                               &                             & $J/\psi$ & $-0.35(6)$  
                                               & $-12(3)$             &                                  
                                               \\ \hline
 \multirow{2}{*}{\cite{Liu:2008rza}}           &                             & $\eta_c$ & $-0.18(9)$  
 & $-9.7(1.2)$          & $14.7(4.8)$                      \\
                                               &                             & $J/\psi$ & $-0.40(80)$ 
                                               & $-12(18)$            & $-100(80)$                       
                                               \\ \hline\hline
                                               &                             & $\,\beta_{J/\psi}~[{\rm GeV}^{-3}]\,$ 
                                               & &                  &                                  
                                               \\
 \multirow{2}{*}{\cite{Sibirtsev:2005ex}}      &                             & $2$      & $-0.37$     
 & $-16.5$              &                                  \\ 
                                               &                             & $0.24$   & $-0.05$     
                                               & $-2.0$               &                                  
                                               
\\ \hline\hline
\end{tabular}
%\end{ruledtabular}
\caption{Summary of the estimates of the low-energy constants $c_0$ and $d_m$ 
discussed in the text. In the upper panel, estimates  from lattice 
determinations of the $S-$wave scattering lengths $a_0$, and in the lower panel, estimates from 
the model in Ref.~\cite{Sibirtsev:2005ex} with the $J/\Psi$ polarizability $\beta$ as in that reference (upper value)
and taken from our fit of the potential in Eq.~\eqref{fpol} (lower value). In quenched 
lattice simulations the value of $d_m$ can not be determined. The values for 
the scattering lengths for the Ref.~\cite{Liu:2008rza} entry correspond to the extrapolation to the 
physical point of our own fit, see Fig~\ref{fqcdslfit}. The scattering lengths for the $J/\psi$ 
channels correspond always to spin averaged values.}
\label{lect}
\end{table}

The $S$-wave quarkonium-nucleon scattering lengths have been studied in the lattice using 
L\"uscher's phase-shift formula in the quenched approximation in Refs.~\cite{Yokokawa:2006td,Kawanai:2010ru} 
and in full QCD in Ref.~\cite{Liu:2008rza}.

In Ref.~\cite{Yokokawa:2006td} the scattering lengths of charmonia ($\eta_c$ and $J/\psi$) with light hadrons 
($\pi$, $\rho$ and $N$) were studied in quenched lattice QCD at three different lattice volumes using L\"uscher's 
phase-shift formula. The full Phase Shift Formula (PSF) as well as a Leading Large $L$ (LLE) expansion were 
employed to extract the scattering lengths from the energy shifts of the system with respect to the sum of the quarkonium
and hadron masses. Three different hopping parameters for the light hadrons were used, 
corresponding to three different pion (and nucleon) masses. Nevertheless, no appreciable light-quark mass dependence was 
found for the values of the energy shifts in the quarkonium-nucleon channels. A possible explanation for this behavior, aside from the fact that the 
simulations are carried out in a quenched approximation, can be derived from our result for the scattering length 
in Eq.~\eqref{slf}; at leading order the scattering length has no light-quark mass dependence and the expected 
size of the light-quark mass dependent subleading contributions to the scattering length is much smaller than the 
size of the uncertainties of the lattice simulations. Thus, we compare our leading order scattering length with 
the results extrapolated in to the physical point presented in Ref.~\cite{Yokokawa:2006td} and extract the value 
of $c_0$. The results can be found in the first entry of Table~\ref{lect}. Note that we have adjusted the sign of 
the scattering lengths to match our convention in Eq.~\eqref{eredef}.

The authors in Ref.~\cite{Kawanai:2010ru} report the values of the scattering length of charmonia ($\eta_c$ and 
$J/\psi$) with nucleons in the second entry of Table~\ref{lect} with an error of the order of $25$\%~. The 
uncertainties are in this case also larger than the expected size of the subleading contributions to the 
scattering length, thus only the value of $c_0$ can be estimated, which may explain why in 
Ref.~\cite{Kawanai:2010ru} also no dependence on the light-quark mass of the scattering lengths was observed. Using the 
lowest pion mass results in that reference, $m_{\pi}=0.64$~GeV ($m_N=1.43$~GeV), we obtain the values 
for $c_0$ presented in Table~\ref{lect}. A value of $1$~fm for the effective range was also reported in 
Ref.~\cite{Kawanai:2010ru} for both channels with $\sim 50$\% uncertainty. Using our leading order expression 
for the effective range we arrive at the estimates $d_{1}+d_{2}=13~{\rm GeV}^{-4}$ for the $\eta_c$ channel and 
$d_{1}+d_{2}=26~{\rm GeV}^{-4}$ for the $J/\psi$ channel with about $60$\% uncertainty.

The scattering lengths of charmonia ($\eta_c$ and $J/\psi$) with light hadrons ($\pi$, $\rho$ and $N$) in full QCD 
were computed in Ref.~\cite{Liu:2008rza}. The Fermilab formulation was used for charm quarks, domain-wall fermions 
for the light-quarks and staggered sea quarks. Four different light-quark masses were used. We fit our expression 
up to NLO for the scattering length of Eq.~\eqref{slf} to the lattice data for the $\eta_c$-N and $J/\psi$-N 
channels. The fit is obtained by minimizing a $\chi^2$ distribution. 

\begin{figure}[ht]
\begin{tabular}{cc}
\includegraphics[width=0.45\textwidth]{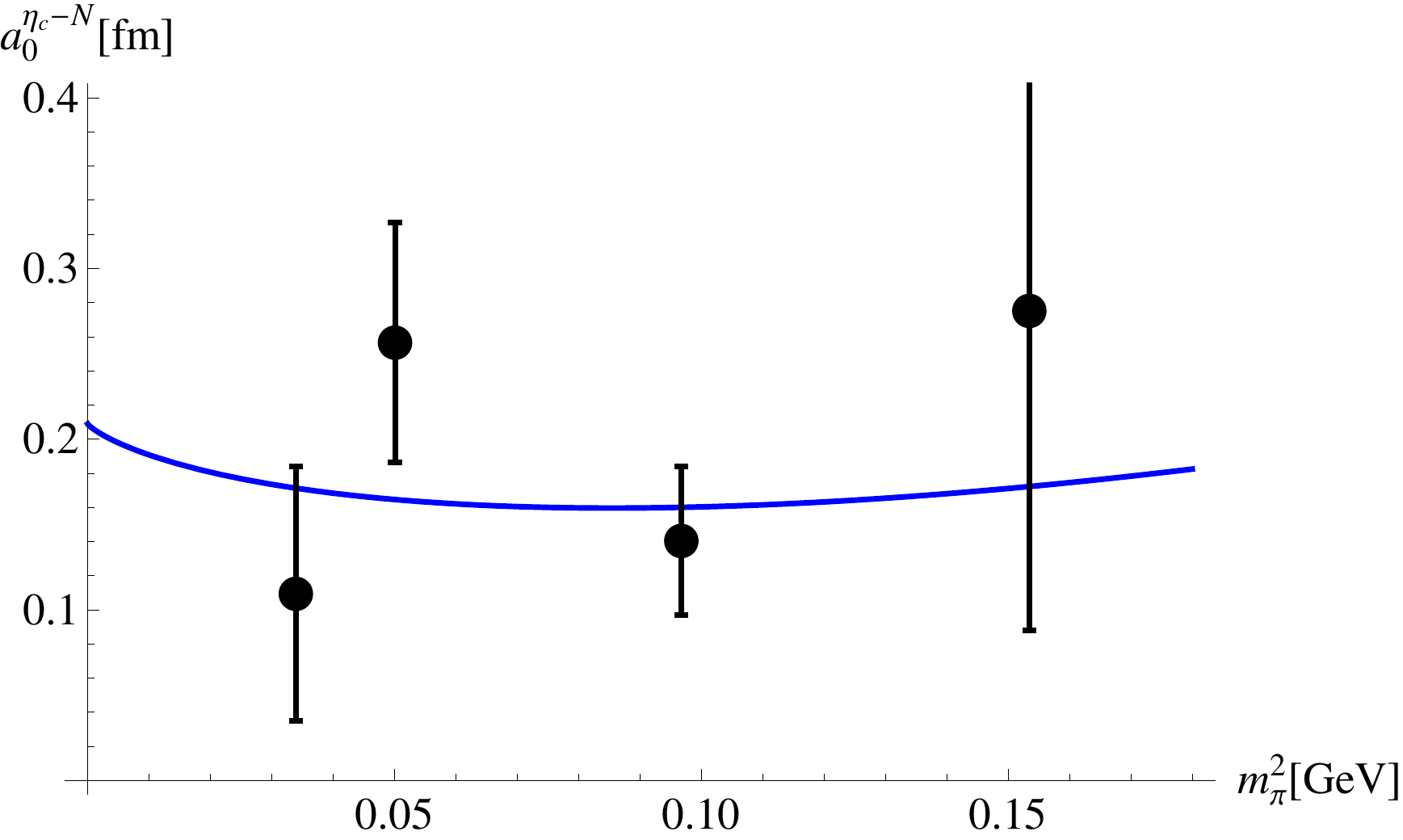} & \includegraphics[width=0.45\textwidth]{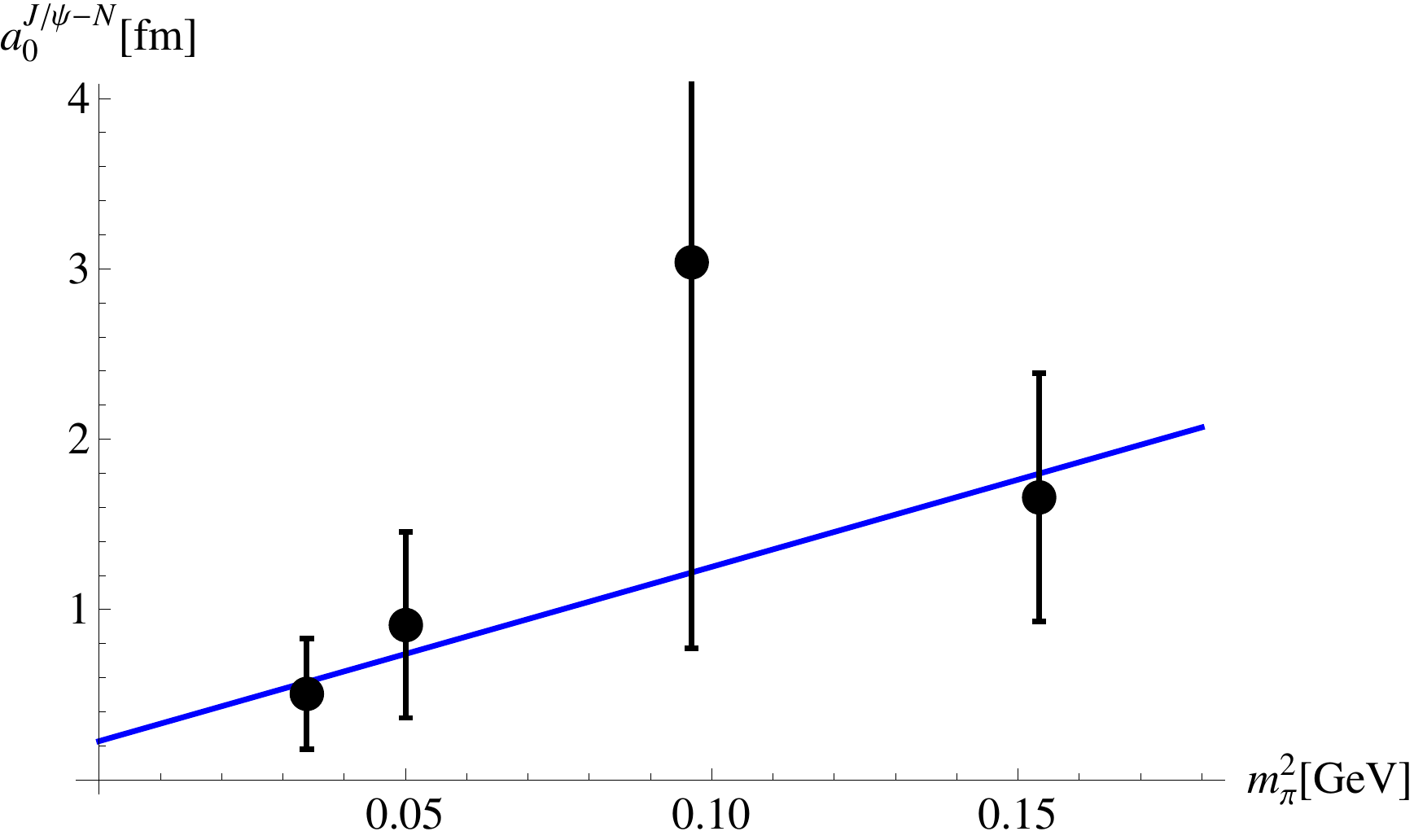} \\
(a)& (b) \\
\end{tabular}
\caption{Fits of the NLO scattering lengths (blue line) as a function of the light-quark mass to the lattice data 
from Ref.~\cite{Liu:2008rza} (black dots). Left- and right-hand panels correspond to the $\eta_c$-N and spin 
averaged $J/\psi$-N channels respectively.}
\label{fqcdslfit}
\end{figure}

At NLO the scattering lengths receive contributions from the light-quark mass dependence of the quarkonium and 
nucleon masses. For the nucleon mass we have used the parameters from fit I of Ref.~\cite{Procura:2003ig}, 
$\hat{m}_N=0.891(4)$~GeV, $c_1=-0.79(5)$~GeV$^{-1}$. We use the values of $\beta$ from Eq.~\eqref{fpol}. The 
quarkonia bare masses are fixed by imposing that Eq.~\ref{mphiq} reproduces the PDG values \cite{Olive:2016xmw} 
($m_{\eta_c}=2.9834$~GeV, $m_{J/\psi}=3.0969$~GeV) at the physical pion mass ($m_{\pi}=0.135$~GeV). We plot the 
fits in Fig.~\ref{fqcdslfit}. The values we obtain are in the third entry of Table~\ref{lect}. The uncertainties 
are estimated as the range of values of a parameter that keeps $\chi^2_{d.o.f}<1$ while keeping the other fixed. 
We can see that the data does not allow to meaningfully constrain the value of $d_m$.

The contact interaction $c_0$ can also be estimated using the model calculation of Ref.~\cite{Sibirtsev:2005ex}. In this 
reference the low-energy interaction of the $J/\psi$ with nucleons was estimated assuming the multipole expansion 
for quarkonium interaction with soft gluons and the soft gluon coupling to the nucleon was determined by the 
anomaly in the trace of the QCD energy-momentum tensor up to an unknown constant $C\geq1$. If we identify the scattering 
amplitude of Ref.~\cite{Sibirtsev:2005ex} (with a nonrelativistic normalization) to our leading order amplitude we obtain
the following estimate for the contact interaction
\be
c_0=-\frac{4\pi^2}{9}\left(1+C\right)\beta_{J/\psi} m_N\,.\label{sivvol}
\ee 
In the lower panel of Table~\ref{lect} we give the value of $c_0$ using Eq.~\eqref{sivvol} with $C=1$ as 
suggested in Ref.~\cite{Sibirtsev:2005ex}. We use two values for the $J/\psi$ polarizability, the first is 
the one used in Ref.~\cite{Sibirtsev:2005ex}, and the second the one obtained in our fit of the potential in 
Eq.~\eqref{fpol}. It should be noted that the quarkonium polarizability in Eq.~\eqref{sivvol} cannot be straightforwardly
identified with the one appearing in Eq.~\eqref{mcm}. In general, the quarkonium polarizability depends on the momentum, for 
distances larger than the inverse of intrinsic energy scale of the quarkonium, i.e. the binding energy, the 
polarizability can be approximated by the zero momentum case (usually called static polarizability in 
electromagnetic systems), which is the case assumed for the polarizability in Eq.~\eqref{sivvol} and fitted in Sec.~\ref{cnpla}. However, at 
shorter distances, such as in the case considered in Ref.~\cite{Sibirtsev:2005ex}, the momentum dependence may 
produce values of the polarizability different from the one fitted in Sec.~\ref{cnpla}.

It is interesting to compare the results, in Table~\ref{lect}, for $c_0$ and $d_m$ obtained from the comparison with 
the different lattice references. There is a moderately large variation in the values obtained for $c_0$, nevertheless 
the larger values also have larger uncertainties. Due to these uncertainties the different values for $c_0$ are not in 
strong contradiction. From naive dimensional analysis we expect $c_0\sim \Lambda^{-2}_{\chi}\sim 1.2$~GeV$^{-2}$, which 
is compatible, within the uncertainties, with the values in Table~\ref{lect}, albeit these lay in the upper range of the 
natural size. The natural size for $d_{1}+d_{2}\sim 3~{\rm GeV}^{-4}$ is smaller than our only estimate from a lattice 
determination of the effective range, this however carry even larger uncertainties than the scattering length ones. If 
values for the low-energy constants exceeding by an order of magnitude the dimensional analysis estimates are confirmed 
by future more precise lattice studies, a different power counting should be adopted. Enhanced contact interactions can 
be accommodated using the PDS scheme of Refs.~\cite{Kaplan:1998we,Kaplan:1998tg} or by the explicit introduction of 
nonscattering quarkonium-nucleon states as degrees of freedom as in Refs.~\cite{Soto:2007pg,Soto:2009xy}.

\section{Comparison with the HAL QCD method potential}\label{cnpla}

The $\eta_c$- and $J/\psi$-nucleon potential has been calculated in the HAL QCD method. In this approach the 
potential is computed in the lattice from the equal-time 
Bethe-Salpeter amplitude through the effective Schr\"odinger equation. The lattice simulations were performed in 
quenched \cite{Kawanai:2010ru} and unquenched \cite{Kawanai:2010ev,Kawanai:2011zz}. We will focus in the 
unquenched results since those are the ones that most directly correspond to our potential, however the authors in 
Ref.~\cite{Kawanai:2010ev} report that there is neither quantitative or qualitative differences between the 
quenched and unquenched results within statistical errors.

The unquenched simulations were carried out using $2+1$-flavor gauge configurations generated by the PACS-CS 
Collaboration on lattices of size $32^3\times64$ with Iwasaki gauge action at $\beta=1.9$, 
which correspond to a lattice spacing of $a^{-1}\approx 2.18$~GeV, and the nonperturbatively $\mathcal{O}(a)$ 
improved Wilson fermions with $c_{SW}=1.715$. The lightest quark masses correspond to $m_{\pi}=0.41$~GeV 
($m_N=1.2$~GeV) and the charm mass corresponds to $m_{\eta_c}=2.99$~GeV and $m_{J/\psi}=3.10$~GeV.

\begin{figure}[ht]
\centerline{\includegraphics[width=.6\textwidth]{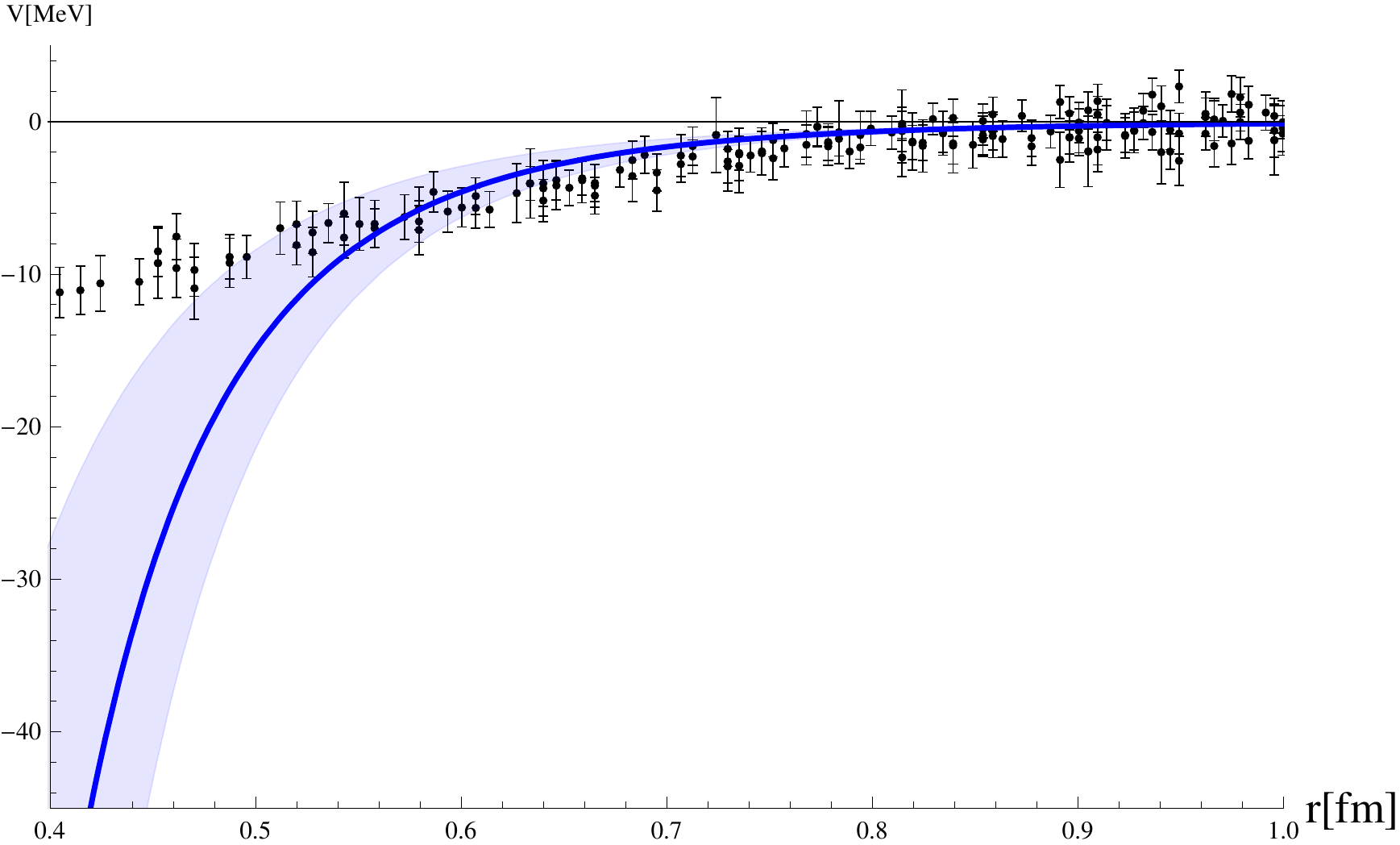}}
\caption{Comparison of the $\eta_c$-nucleon potential obtained with unquenched simulations 
from Ref.~\cite{Kawanai:2010ev} (black dots) with our result for the nonlocal part of the potential in 
Eq.~\eqref{Vnloc} (blue line). The light blue band displays the theoretical uncertainty due higher order 
contributions suppressed by a factor of order $(r\Lambda_{\chi})^{-1}$ with respect to the 
leading terms.}
\label{etacNpc}
\end{figure}

\begin{figure}[ht]
\centerline{\includegraphics[width=.6\textwidth]{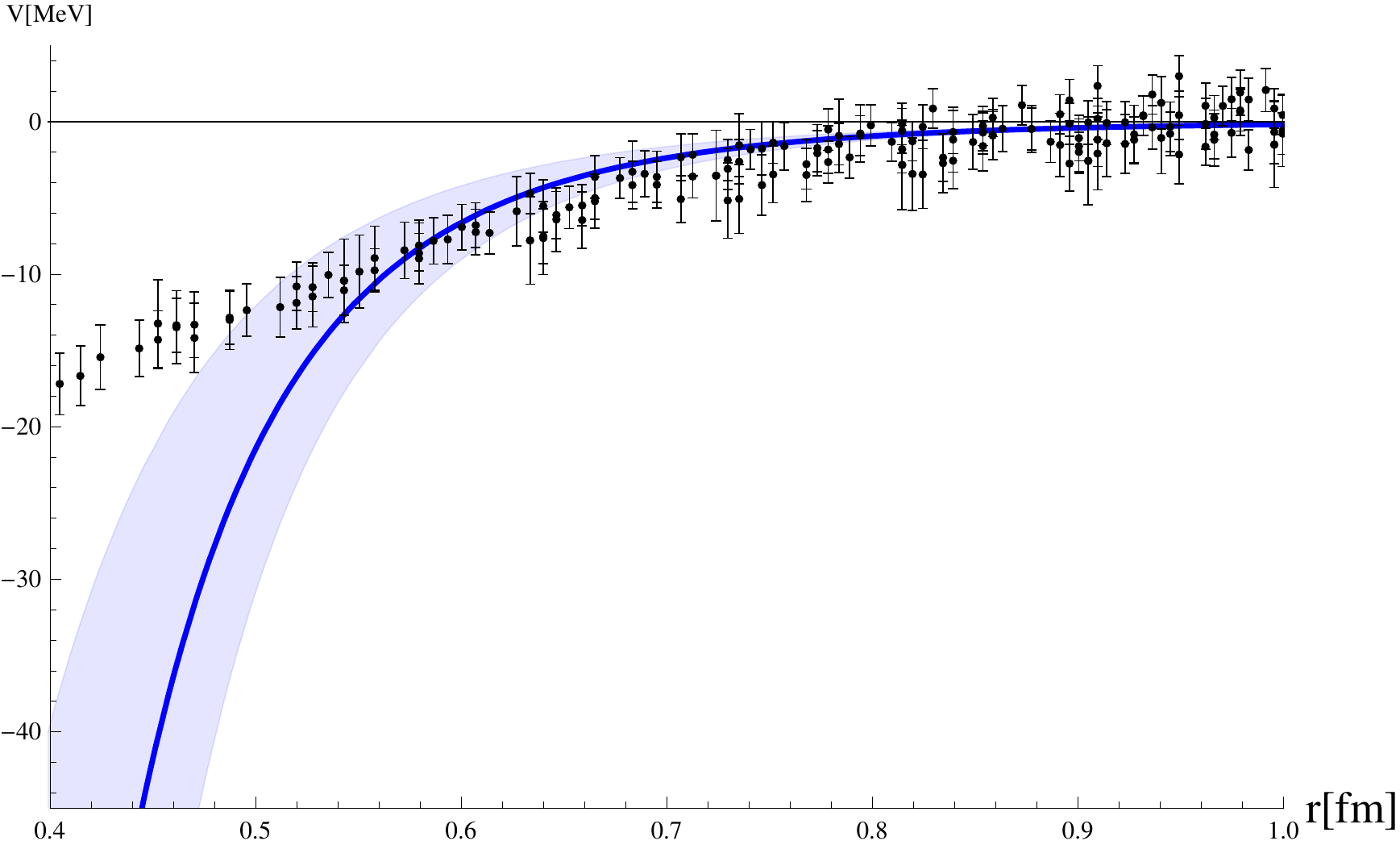}}
\caption{Comparison of the $J/\psi$-nucleon potential obtained with unquenched simulations from Ref.~\cite{Kawanai:2010ev} 
(black dots) with our result for the nonlocal part of the potential in Eq.~\eqref{Vnloc} (blue line). The light 
blue band displays the theoretical uncertainty due higher order contributions suppressed by a factor of order $(r
\Lambda_{\chi})^{-1}$ with respect to the leading terms.}
\label{jpsiNpc}
\end{figure}

The lattice data extends to fairly short distances of about $\sim0.2$~fm, which correspond to momentum transfers 
far beyond the applicability of our EFT approach, so we will compare our potential in Eq.~\eqref{Vnloc} only with data at $r\gtrsim 0.4$~fm 
corresponding to momentum transfers $|\bm{k}|\lesssim 0.5$~GeV. The value of the axial coupling and the pion decay 
constants in the chiral limit are taken as $g_A=1.2$ and $F=0.0862$~GeV from Ref.~\cite{Hemmert:2003cb} and 
\cite{Baron:2009wt} respectively. We use the same value of the pion mass $m_{\pi}=0.410$~GeV as the lattice 
simulations.

We use the expressions of the pion-quarkonium couplings in Eq.~\eqref{mcm} in the expression of the 
potential in Eq.~\eqref{Vnloc} and arrive to an expression with the polarizability $\beta$ as the only free parameter.
We fit the potentials to the lattice data and obtain the value for the polarizabilities. The fit is obtained by minimizing a $\chi^2$ distribution where to the statistical errors of the lattice we add the theoretical uncertainties. The values of the polarizabilities from the fit vary depending on the range $r$ used, 
but stabilize for ranges with lower $r$ cutoff of the order of $0.50$~fm. We consider the range $0.5<r<1.4$~fm a 
good compromise between using the most lattice data and obtaining a stable fit. The values of the charmonium 
polarizabilities obtained are
\be
\beta_{\eta_c}=0.17~{\rm GeV}^{-3}\,,\quad \chi^2_{d.o.f}=0.7\,,\quad
\beta_{J/\psi}=0.24~{\rm GeV}^{-3}\,,\quad \chi^2_{d.o.f}=0.7\,,\label{fpol}
\ee
The corresponding values of the pion-quarkonium couplings are shown in Table~\ref{qpcv}.
\begin{table}
%\begin{ruledtabular}
\begin{tabular}{c|c|c|c} \hline\hline
                                     & $c_{d0}~[{\rm GeV}^{-3}]$ & $c_{di}~[{\rm GeV}^{-3}]$ & $c_{m}~[{\rm GeV}^{-3}]$ \\ \hline
$\beta_{\eta_c}=0.17~{\rm GeV}^{-3}$ & $-0.83$                   & $-1.71$                   & $-2.24$                  \\
$\beta_{J/\psi}=0.24~{\rm GeV}^{-3}$ & $-1.17$                   & $-2.42$                   & $-3.16$                  \\ \hline\hline
\end{tabular}    
%\end{ruledtabular}
\caption{Values of the pion-quarkonium couplings according to the expressions in Eq.~\eqref{mcm} for the values of the polarizabilities, in Eq.~\eqref{fpol}, obtained from the fit of the potential to the lattice data of Ref.~\cite{Kawanai:2010ev}.}
\label{qpcv}
\end{table}

Note that the $\chi^2_{d.o.f}$ does not include the theoretical uncertainty. In Fig.~\ref{etacNpc} we compare the 
$\eta_c$-nucleon potential obtained with the HAL QCD method with the potential in Eq.~\eqref{Vnloc}. In Fig.~\ref{jpsiNpc} 
we show the analogous comparison with the lattice $J/\psi$-nucleon potential. The light-blue band displays the theoretical 
uncertainty, which we estimate by adding or subtracting contributions suppressed by a factor $(r\Lambda_{\chi})^{-1}$ with 
respect to our potential. The discussion regarding the subleading contributions can be found in Sec.~\ref{pnqeft} in the paragraph
following Eq.~\eqref{swpot}. The values of the polarizability obtained from the fit can be compared to the perturbative 
(pNR)QCD calculation in Refs.~\cite{Voloshin:1979uv,Leutwyler:1980tn,Brambilla:2015rqa}. The values of $\beta$ from the 
fit are reproduced by the perturbative QCD formula for the values $\alpha_s=1.3$ and  $\alpha_s=1.2$ for the $\eta_c$ 
and $J/\psi$ respectively. 

\begin{figure}[ht]
\centerline{\includegraphics[width=.6\textwidth]{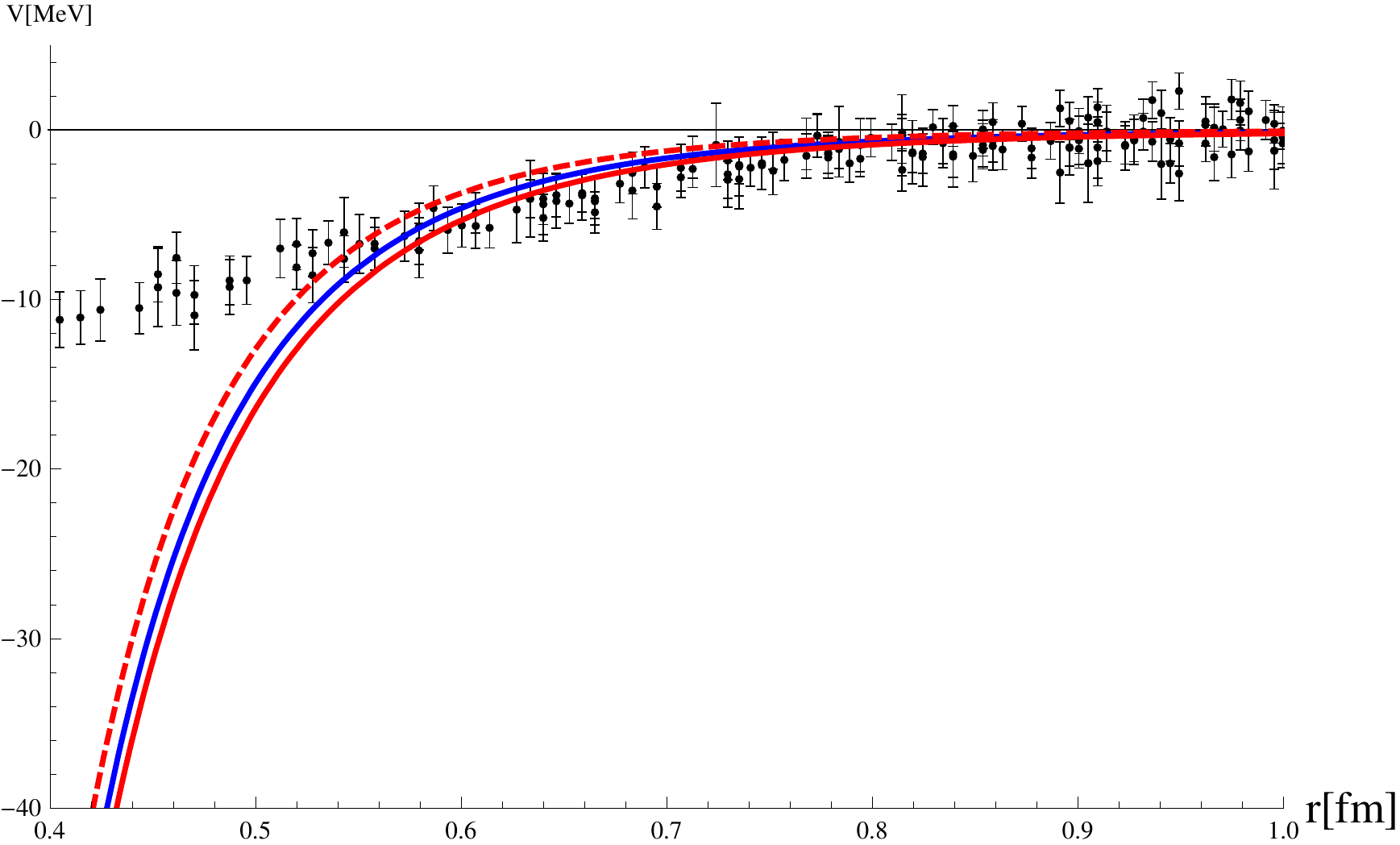}}
\caption{Comparison of the nonlocal part of the $\eta_c$-nucleon potential for different pion masses with the 
results from unquenched simulations from Ref.~\cite{Kawanai:2010ev} (black dots). The blue 
line corresponds to $m_{\pi}=0.410$~GeV, the red line to $m_{\pi}=0.275$~GeV, and the red dashed line to $m_{\pi}
=0.545$~GeV. The differences in the potential for the three different pion masses are small compared to the 
uncertainties of the lattice data.}
\label{etacNpcm}
\end{figure}

In Fig.~\ref{etacNpcm} we plot our potential for three pion masses, $m_{\pi}=0.410$~GeV, $m_{\pi}=0.275$~GeV and 
$m_{\pi}=0.545$~GeV together with the lattice data. We find that the variation of the potential is small compared 
to the uncertainties of the lattice data. In Refs.~\cite{Kawanai:2010ru,Kawanai:2010ev} it was noted that the 
unquenched results for the potential showed small variations with the light-quark mass which is consistent with 
our results. In Ref.~\cite{Sugiura:2017vks} the progress on an improved determination of the quarkonium-nucleon 
potential, based also  HAL QCD method, in which ground state saturation is more easily achieved, was reported. No 
significant difference with Refs.~\cite{Kawanai:2010ru,Kawanai:2010ev} was noted.

\section{Conclusions}\label{conc}

We have developed an EFT description of the $S$-wave quarkonium-nucleon system and obtained the quarkonium-nucleon 
potential, scattering length and effective range up to $\mathcal{O}(m^3_{\pi}/\Lambda^3_{\chi})$ accuracy. We have 
compared our results with lattice QCD studies of quarkonium-nucleon system. In Sec.~\ref{nqeft} we have constructed 
the quarkonium-nucleon EFT (QNEFT), at energies of order $E\sim m_{\pi}$, much below $\Lambda_{\chi}$ by writing the 
most general Lagrangian with one quarkonium, one nucleon and pions consistent with chiral symmetry, $C$ and $P$ 
invariance and rotational symmetry (and Lorentz invariance in the pion sector). We have adopted a power counting 
given by dimensional analysis, equivalent to Weinberg's power counting in nucleon-nucleon EFTs. We have noted that 
the quarkonium-nucleon dynamics occurs at a lower energy scale $E\sim m^2_{\pi}/\Lambda_{\chi}$. We have integrated 
out the $E\sim m_{\pi}$ modes and matched QNEFT to a lower energy EFT which we called potential QNEFT (pQNEFT) in 
Sec.~\ref{pnqeft}. In this EFT there are no longer dynamical pion fields and its effects are taken into account 
through potential interactions and redefinitions of the low-energy constants. In Sec.~\ref{erep} we have computed 
the quarkonium-nucleon $S$-wave scattering amplitude in pQNEFT and obtained the expressions for the scattering 
length and effective range including the light-quark mass dependence.

In Sec.~\ref{cnplb} we have compared our result for the scattering length with the lattice QCD determinations of 
Refs.~\cite{Yokokawa:2006td,Kawanai:2010ru,Liu:2008rza} and extracted the value of the leading quarkonium-nucleon 
contact term coefficient $c_0$. There is significant dispersion on the values of $c_0$ obtained from the different 
sources of lattice data, however results are consistent within uncertainties. The most accurate determinations of 
$c_0$ point to a value of $c_0\approx 8$GeV$^{-2}$ which is on the upper band of the size range expected from 
dimensional analysis. Therefore, Weinberg's power counting seems to be consistent with the lattice data but 
alternative power counting schemes allowing for enhanced contact terms are not ruled out. 

We have compared our results for the quarkonium-nucleon potential obtained from the HAL QCD method in 
Sec.~\ref{cnpla}. Fitting our potential to the data of Refs.~\cite{Kawanai:2010ru,Kawanai:2010ev} we determine 
the value of the polarizability of the $J/\Psi$ and the $\eta_c$, $\beta_{J/\psi}=0.24~{\rm GeV}^{-3}$ and 
$\beta_{\eta_c}=0.17~{\rm GeV}^{-3}$ respectively. Both in the comparisons of the potential and the 
scattering length we have found that the light-quark mass dependence is no appreciable with the current 
accuracy of the data.

Let us now comment on the implication of the results of Sec.~\ref{cnplb} in the possibility of the existence of 
quarkonium-nucleon bound states. In order to poles to appear in the scattering amplitude one would need 
to resum the scattering amplitude in Eqs.\eqref{pefta0}-\eqref{pefta3} with respect to the LO 
result. Up to N$^2$LO the resumed amplitude is equivalent to the ERE and the nonlocal 
two-pion exchange potential appears only at N$^3$LO. Therefore, the position of the pole is given at 
leading order is $ip_b\sim -\frac{2\pi}{\mu c_0}$, which for our best estimates of $c_0$ takes the value 
$ip_b\sim 1$GeV far beyond the range of applicability of our EFT. Finally, it is interesting to compare 
with the binding energies in the quarkonium-nucleon system obtained in Ref.~\cite{Alberti:2016dru}. In this 
work the heavy-quark-antiquark static potentials with a background 
hadron were obtained in a lattice QCD simulation at a pion mass of about $223$~MeV. The authors report a 
binding energy between nucleon and $1S$-charmonium of $-2.4$~MeV. This corresponds to relative momentum 
$p_b\sim 58$~MeV, much smaller than our previous estimate and corresponds to larger values of $c_0$ than 
any found in Table~\ref{lect}. 

\bigskip
{\bf Acknowledgments}
\bigskip

We thank Taichi Kawanai for providing us access to the lattice data for the charmonium-nucleon potential 
and Nora Brambilla and Antonio Vairo for reading the manuscript and suggestions.  
J.T.C has been supported by the DFG and the NSFC through funds provided to the Sino-German CRC 110 
``Symmetries and the Emergence of Structure in QCD'' and by the DFG cluster of excellence ``Origin and 
Structure of the Universe''  and by the Spanish MINECO's grants Nos.\ FPA2014-55613-P, FPA2017-86989-P 
and SEV-2016-0588. The work of G.K. was partially financed by Conselho Nacional de Desenvolvimento 
Cient\'{\i}fico e  Tecnol\'ogico - CNPq, Grant No. 305894/2009-9 and No. 464898/2014-5 (INCT F\'{\i}sica 
Nuclear e Applica\c{c}\~oes), and Fun\-da\-\c{c}\~ao de Amparo \`a Pesquisa do Estado 
de S\~ao Paulo Grant No. 2013/01907-0.

\appendix

\section{Potential in coordinate space}\label{appa}

The two-pion exchange diagram (d) in Fig.~\ref{nqsd} contains both local and nonlocal (potential) 
contributions. A dispersive representation is a natural form to split the short- and long-range contributions 
in momentum space, as well as a convenient way to obtain a coordinate space representation of a potential. 
Let $\bm{p}$ and $\bm{p}^{\prime}$ be the c.m. momenta of the incoming and outgoing nucleon and
${\bm k}=\bm{p}-\bm{p}^{\prime}$ the momentum transfer. The amplitude corresponding to diagram (d) in 
Fig.~\ref{nqsd} is given by
\begin{align}
\mathcal{A}_{\text{(d)}} =& - \frac{3g^2_Am^3_{\pi}}{32\pi F^2}
\left\{
2c_{di}\left(1+\frac{\bm{k}^2}{4m^2_{\pi}}\right)-c_m
+ \left[c_{di}\left(1+\frac{\bm{k}^2}{2m^2_{\pi}}\right)-c_m\right]
\right. \nn \\
&\left. \times
\sqrt{\frac{m^2_\pi}{\bm{k}^2}} \left(1+\frac{\bm{k}^2}{2m^2_{\pi}}\right)
\arctan\sqrt{\frac{\bm{k}^2}{4m^2_{\pi}}}\right\}\,.
\label{nqqbnnlo}
\end{align}
Since the nonanalytic piece of the amplitude diverges as $\bm{k}^3$ for large $\bm{k}$, we need a 
twice-subtracted dispersive representation. In Ref.~\cite{Brambilla:2015rqa} such a method was 
employed to extract the long-distance part of the low-energy $\eta_b-\eta_b$ interaction.

We set the subtraction points at $0$. Using partial fractioning we arrive at
\begin{align}
\mathcal{A}(s) =& \mathcal{A}(0)+\left.s\,\frac{d\mathcal{A}}{ds}\right|_{s=0}
- \frac{s}{\pi}\int^{\infty}_{4m^2_{\pi}}\frac{\text{Im} \mathcal{A}(s^{\prime})}{(s^{\prime})^2}ds^{\prime}
- \frac{1}{\pi}\int^{\infty}_{4m^2_{\pi}}\frac{\text{Im} \mathcal{A}(s^{\prime})}{s^{\prime}}ds^{\prime} 
\nn \\
&+\frac{1}{\pi}\int^{\infty}_{4m^2_{\pi}}\frac{\text{Im} \mathcal{A}(s^{\prime})}
{s^{\prime}-s-i\epsilon}ds^{\prime}\,,
\end{align}
with $s=-\bm{k}^2$. Setting $s^{\prime}=\mu^2+i\epsilon$ and using that
\be
\text{Im}\left[\frac{1}{\epsilon-i\mu}\arctan\left(\frac{\epsilon-i\mu}{2m_{\pi}}\right)\right]
= \frac{\pi}{2\mu}\theta(\mu-2m_{\pi}),
\ee
we obtain
\begin{align}
- \mathcal{A}_{\text{(d)}}=&\frac{3g^2_Am^3_{\pi}}{64\pi F^2}
\left(5c_{di}-3c_{m}+\sigma_0\right)
+ \frac{g^2_A m_{\pi}}{256\pi F^2} \left(23c_{di}-5c_{m}+\sigma_1\right)\bm{k^2} \nonumber\\
& + \frac{3g^2_Am^3_{\pi}}{64\pi F^2}\int^{\infty}_{1}dx \, 
\frac{\left[c_{di}\left(1-2x^2\right)-c_m\right]\left(1-2x^2\right)}{x^2+\frac{\bm{k}^2}{4m^2_{\pi}}}
\,,
\label{dsplited}
\end{align}
where 
\begin{align}
&\sigma_0 = -4\int^{\infty}_{1}dx \, \frac{1}{x} \left[c_{di}\left(1-2x^2\right)-c_m\right]
\left(1-2x^2\right)\,, 
\\
&\sigma_1 = \frac{16}{3}\int^{\infty}_{1}dx \, 
\frac{1}{x^4} \left[c_{di}\left(1-2x^2\right)-c_m\right] \left(1-2x^2\right)\,,
\end{align}
are the subtraction constants that can be reabsorbed in the low-energy constants. The nonlocal part 
of the amplitude is then obtained by subtracting the contact terms from the original amplitude 
given in Eq.~(\ref{nqqbnnlo}), namely:
\begin{align}
{\mathcal A}^{\text{long}}_{\text{(d)}}({\bm k}^2) =& - \frac{3g^2_Am^3_{\pi}}{32\pi F^2}\left\{\left(
c_{di}-c_m 
+ c_{di}\frac{\bm{k}^2}{2m^2_{\pi}}\right)\left[\sqrt{\frac{m^2_{\pi}}{\bm{k}^2}}
\left(1+\frac{\bm{k}^2}{2m^2_{\pi}}\right)\arctan\sqrt{\frac{\bm{k}^2}{4m^2_{\pi}}}
-\frac{1}{2}\right] \right.\nonumber\\
&\left.-(c_{di}-c_m)\frac{5\bm{k}^2}{24m^2_{\pi}}\right\}\,.
\label{V-mom}
\end{align}
The long-range part of the quarkonium-nucleon potential in momentum space is then given 
by $\widetilde{V}({\bm k}^2) = - {\mathcal A}^{\text{long}}_{\text{(d)}}({\bm k}^2$).

The first two pieces in Eq.~(\ref{dsplited}) are matched into $C_0$ and $D_1$ respectively, and 
the last term is used to obtain the potential in coordinate space through
\begin{align}
V(r) &= \int\frac{d^3\bm{k}}{(2\pi)^3} \, e^{i\bm{k}\cdot\bm{r}} \; \widetilde{V}(\bm{k}) \nn \\
&=\frac{3g^2_Am^5_{\pi}}{64\pi^2 F^2 r}\int^{\infty}_{1}dx\,
e^{-2m_{\pi}r x}\left[c_{di}\left(1-2x^2\right)-c_m\right]\left(1 - 2x^2\right)\,.
\end{align}
Performing the integral in $x$ we arrive at the result in Eq.~(\ref{Vnloc}).

\bibliographystyle{apsrev4-1}
\bibliography{qneftbib}

\end{document}